 \documentclass[preprint2]{aastex}

\shorttitle{NS Kicks from Asymmetries}
\shortauthors{Fryer}

\begin{document}

\title{Neutron Star Kicks from Asymmetric Collapse}


\author{Chris L. Fryer} 

\affil{Theoretical Astrophysics, Los Alamos National Laboratory, Los
Alamos, NM 87545; Physics Department, The University of Arizona,
Tucson, AZ 85721}

\begin{abstract}
  
  Many neutron stars are observed to be moving with spatial
  velocities, in excess of 500\,km\,s$^{-1}$.  A number of mechanisms
  have been proposed to give neutron stars these high velocities.  One
  of the leading classes of models proposed invokes asymmetries in the
  core of a massive star just prior to collapse.  These asymmetries
  grow during the collapse, causing the resultant supernova to also be
  asymmetric.  As the ejecta is launched, it pushes off (or ``kicks'')
  the newly formed neutron star.  This paper presents the first
  3-dimensional supernova simulations of this process.  The ejecta is
  not the only matter that kicks the newly-formed neutron star.
  Neutrinos also carry away momentum and the asymmetric collapse leads
  also to asymmetries in the neutrinos.  However, the neutrino
  asymmetries tend to damp out the neutron star motions and even the
  most extreme asymmetric collapses presented here do not produce
  final neutron star velocities above 200\,km\,s$^{-1}$.

\end{abstract}


\section{Introduction}

A neutron star is born when the core of a massive star collapses to
nuclear densities during the last stages of the star's life.  It is
during, or just after, the launch of a supernova explosion associated
with neutron star birth that it is believed that the nascent neutron
stars receive the fast spins and large magnetic fields that allow them
to be observed as pulsars.  The inferred velocities from individual
pulsars (e.g. Guitar Nebula Pulsar - Cordes, Romani, \& Lundgren 1993)
and pulsar velocity distributions (Lyne \& Lorimer 1994; Lorimer,
Bailes, \& Harrison 1997; Fryer, Burrows, \& Benz 1998; Cordes \&
Chernoff 1998; Brisken et al. 2003) imply that neutron stars also
receive a kick, most likely at birth.  In addition, specific neutron
star and black hole binaries are best explained by assuming neutron
stars receive large kicks at birth (e.g. Fryer \& Kalogera 1997;
Kramer 1998; Tauris et al. 1999; Wex, Kalogera, \& Kramer 2000;
Mirabel et al. 2002).  Any supernova mechanism must not only produce
explosions, but also must expain features of nascent neutron stars,
such as spin, magnetic fields and high velocities.

Of all of these, theorists have been most hard pressed to explain the
neutron star velocities.  Most mechanisms by which these kicks might
be produced require extreme magnetic fields (see Lai, Chernoff, \&
Cordes 2001 for a review).  One of the leading classes of models for
kick production which avoids these strong magnetic fields invokes
global hydrodynamic perturbations in the stellar core prior to
collapse.  These perturbations are driven during the last few weeks of
a star's life.  At the end of its life, a star is composed of an iron
core covered by a succession of composition layers (from silicon down
to hydrogen) produced by a series of nuclear burning stages where the
ashes of each burning stage is the fuel for the next.  The final
stages of a star's life are marked by explosive burning in the oxygen
and silicon layers above the iron core (Bazan \& Arnett 1998).  The
large scale convection driven by the burning creates density
variations in these oxygen and silicon shells. Such asphericities,
especially if dominated by low order (l=1,2) modes, cause asymmetries
during the collapse, bounce and explosion of the stellar interior
(Burrows \& Hayes 1996).  It is possible that the violent convection
in shell burning can also excite unstable modes in the star's iron
core, producing asymmetries in the iron core itself, increasing the
level of asymmetry in the explosion (Lai \& Goldreich 2000).  These 
asymmetries are believed to grow during collapse and bounce and 
ultimately, to produce asymmetric explosions and large neutron star 
velocities.

In this paper, the first 3-dimensional collapse simulations.of this
asymmetric-collapse kick mechanisms are presented.  We compare and
contrast these simulations to the seminal 2-dimensional work by
Burrows \& Hayes (1996) which predicted strong neutron star kicks 
from this mechanism.

\section{Collapse Code and Initial Conditions}

Our collapse calculations use the 3D SNSPH code (Fryer \& Warren 2002;
Warren, Rockefeller, \& Fryer 2003).  This code combines a parallel
Lagrangean hydrodynamics scheme (smooth particle hydrodynamics - SPH) 
with a flux-limited diffusion neutrino transport scheme.  Gravity
is calculated using a tree-based algorithm described in Warren \&
Salmon (1995).  The neutrino transport and equation of state physics
uses the same algorithms described in Herant et al. (1994) and Fryer
(1999).  The Lagrangean nature of SPH is critical for this problem, as
the proto-neutron star is not fixed to any central point and the
resolution must follow this moving proto-neutron star.  The entire
4$\pi$ of the star is modeled in 3-dimensions.

Beyond a neutrino optical depth, $\tau_\nu$, of 0.03, the code assumes
the neutrinos escape the star without further affecting the
composition or hydrodynamics of the star.  But we must still account
for the momentum lost by neutrinos at this boundary.  We assume
particles emit neutrinos isotropically over a hemisphere with a net
momentum emitted set to $0.5 (E_\nu/c) \hat{r}$ where $E_\nu$ is the
energy of the neutrinos, $c$ is the speed of light, and $\hat{r}$ is
the radial direction of the particle.  The momentum lost is taken from
the emitting particle.  In reality, neutrinos will have some beaming
and the true answer for the momentum carried away by neutrinos lies
between $0.5-1.0 (E_\nu/c) \hat{r}$.

For initial conditions, we use as a base model the 15\,M$_\odot$
progenitor ``s15s7b2'' (Woosley \& Weaver).  To this, we have added
the following two perturbations.  To mimic the low mode convection in
the oxygen and silicon layers only, we have run a simulation where the
density in the oxygen and silicon layers only is lowered by 40\% in a
30$^{\circ}$ cone around the z-axis: shell only simulation.  Neutron
star oscillations are mimicked by lowering the density throughout the
collapsing star (including the iron core) by 25\% in a 30$^{\circ}$
cone around the z-axis: core oscillation simulation.  These
perturbations are extreme cases.  Bazan \& Arnett (1998) predict
density variations at the 10-15\% level with higher modes.  Burrows \&
Hayes used a 15\% asymmetry in a 20$^{\circ}$ wedge.  In running test
simulations with perturbations at this 15\% level with a 20$^{\circ}$
wedge, the neutron star kicks did not exceed 30\,km\,s$^{-1}$.

\section{Simulation Results}

The collapse simulations follow the collapse of these modified cores
through bounce until an explosion shock is launched (Figs. 1,2).
After 80\,ms, the newly-born neutron star (central black dot defined
by material with density above $10^{13} {\rm g\, cm^{-3}}$) has moved
$\sim$16km, corresponding to an average velocity of 200\,km\,s$^{-1}$.
But this estimate can be misleading because the neutron star (defined
by this density requirement) is growing during this time,
preferentially accreting along the negative z axis.  So part of this
motion is an artifact of this one-sided growth.  Also, this assumes a
steady motion of the neutron star and, as we shall see below, this
motion is far from steady.  At this time, the convective region is
just beginning to overcome the ram pressure of the infalling star and
is now expanding.

This kick magnitude is much lower than that predicted by Burrows \&
Hayes (1996).  Before we delve into the details and differences of
this simulation with respect to Burrows \& Hayes (1996), let's review
their calculations.  The 2-dimensional simulations by Burrows \& Hayes
(1996) assumed a symmetry along the z-axis where the density
perturbation was placed.  The inner 15\,km of the star was modeled in
1-dimension and was held fixed to a central position.  This means that
the proto-neutron star does not move and the kick can only be
determined by assuming momentum conservation.  In such conditions,
mild asymmetries of 15\% in a 20$^\circ$ cone produced highly
asymmetric supernovae with kicks in excess of 450\,km\,s$^{-1}$.

In the simulations presented here, the initial bounce began even more
asymmetric than those of Burrows \& Hayes (1996) due to the more
extreme initial conditions.  After bounce, the neutron star begins to
move in the negative-z direction.  The convection that drives the
explosion rises from the surface of the neutron star.  As the neutron
star moves, the convection moves with it, weakening the explosion in
the postive-z direction.  The net result is that the explosion is only
mildly asymmetric (Figs. 1,2).

Not only is the explosion less asymmetric and kick magnitude weaker than
that of Burrows \& Hayes (1996), but even the qualitative nature of
the kick is very different.  Instead of accelerating continuously in
the negative z-direction, the velocity of the newly born neutron star
oscillates substantially.  Figure 3 shows the x,y, and z velocities of
the proto-neutron star (again defined as that material with densities
in excess of $10^{13} {\rm g\, cm^{-3}}$) for both simulations as a
function of time since collapse.  Note that in the core oscillation
model, decreasing the mass in the core caused the collapse to take
20\% longer.  However, though its density profile extends throughout
the entire core, the velocities of this simulation with a 25\% density
decrease are less than the shell-only simulation with its 40\% density
perturbation.  It appears that it is the density perturbation in the
oxygen/silicon shells that matters most for kicks.  But in both, the
same oscillatory behavior occurs.

Let's gain a better understanding of this oscillating motion.  During
collapse, the condensing core moves along the path of least
resistance in the positive z-direction where we decreased the
density.  The bounce shock is strongest in this direction, and
momentum conservation argues that the neutron star should be driven in
the negative z-direction, just as predicted by the 2-dimensional
momentum conservation arguments (Burrows \& Hayes 1996).

Because the ejecta is roughly symmetric, the momentum carried away by
neutrinos dominates the momentum loss from the core.  As the neutron
star moves through the convective region (Fig. 2), the down-flows, and
hence neutrino emission, are strongest along the leading edge of the
neutron star.  These neutrinos exert a force on the neutron star as
they carry momentum out of the star, causing the neutron star to
switch directions once again.  The neutrino forces gradually damp the
neutron star's motion.  As the explosion evolves, the rapid shaking of
the neutron star ends, leaving behind a slowly moving
($<200$\,km\,s$^{-1}$) neutron star.

\section{Implications}

The 3-dimensional simulations presented here {\it did} find that
density perturbations can drive a neutron star to velocities in excess
of 700\,km\,s$^{-1}$.  However, neutrino emission damps out these
velocities by the time the star explodes.  The simulations did not
produce strong ($> 500$\,km\,s$^{-1}$) neutron star kicks even with
the very extereme density perturbations in the pre-collapse star
assumed here.  Since our 15\,M$_\odot$ is fairly representative of
stellar progenitors above 12-13\,M$_\odot$, our simulations suggest
something is missing in the standard, neutrino-driven mechanism for
both type Ib/c and type II supernovae alike.  The progenitors
of type Ib/c and type II supernova differ mostly by the amount of
hydrogen, or lack thereof in the envelope.  The cores of type Ib/c and
II supernovae should be very similar, so our results will stand for 
both types of supernovae.

Hence, we can place severe constraints on the standard,
neutrino-driven supernova engine: either 1) the explosion must occur
very quickly before neutrino emission from accretion flows can damp
the neutron star motion, 2) the explosion must occur late, allowing
the convective flows to merge into single modes (Herant 1995; Scheck
et al. 2003), 3) magnetic fields are indeed high ($> 10^{15}$G) at
least in local patches early on in the proto-neutron star (Arras \&
Lai 1999; Socrates et al. 2003).  Determining which of these
mechanisms, if any, will work can teach us much about the supernova
engine.

Roughly half of all neutron stars do receive a strong kick (Fryer,
Burrows, \& Benz 1998; Arzoumanian, Chernoff, \& Cordes 2002).  If
collapse asymmetries are to produce the kick, we need the explosion to
occur early.  One way to achieve fast explosions is to rely upon low
mass progenitors.  Low mass progenitors (8-12\,M$_\odot$) will not
have a large envelope to prevent the explosion and could explode
quickly, producing strong kicks.  Since they make up half of all
core-collapse events, it may be that these stars produce all the
fast-moving neutron stars.  But the progenitors for such systems are
difficult to produce, and until a realistic progenitor is produced, it
is difficult to test this scenario.

The supernova explosion mechanism must also explain the explosion
asymmetries observed in supernovae (see Akiyama et al. 2003 for a
review).  As with neutron star kicks, our 3-dimensional simulations
found that the level of ejecta asymmetry was much less than that
predicted by 2-dimensional simulations.  Our asymmetric collapse
simulations produce ejecta with $v_{\rm pole}/v_{\rm eq}$ ratios not
larger than 1.2, much lower than those predicted by the observations.  
But low mass progenitors or the merger of convective modes (Scheck et 
al. 2003) may produce sufficient asymmetries to explain the observations.

The asymmetry in the neutrino emission will dominate the gravitational
wave signal (Burrows \& Hayes 1996) for these density perturbations.
Using the formulism set by M\"uller \& Janka (1997), we have
calculated the gravitational wave signal from neutrino asymmetries
both above (positive z axis) and below (negative z axis) the density
perturbation (Fig. 4).  Note that the simulation including
pertubations due to iron core oscillations produces a much stronger
gravitational wave signal, despite the lower density perturbations.
The gravitational wave signal can thus be used to study the motions of
the pre-collapse stellar core (Fryer, Holz, \& Hughes 2003).

\acknowledgements

It is a pleasure to thank T. Janka and L. Wang for extremely helpful
advice on this paper.  This work has been funded by DOE SciDAC grant
number DE-FC02-01ER41176, a LANL-based ASCI grant, and NASA NAG5-9192.

\begin{figure}
\plotone{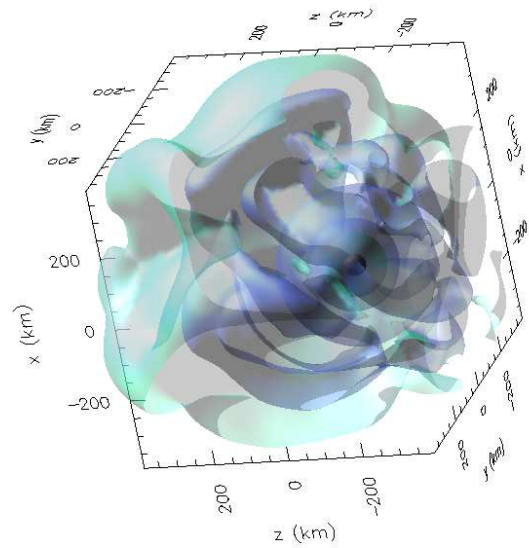}
\caption{3-dimensional view of the exploding core of the shell-only 
  simulation 80\,ms after collapse.  The solid sphere in the center
  marks the nascent neutron star (defined by that material with
  densities above $10^{13} {\rm g\, cm^{-3}}$).  The contours show are
  entropy isosurfaces showing the turbulent convective region just now
  developing into a supernova explosion.  The neutron star is already
  moving off-center, but the explosion itself is only mildly
  asymmetric.}
\end{figure}

\begin{figure}
\plotone{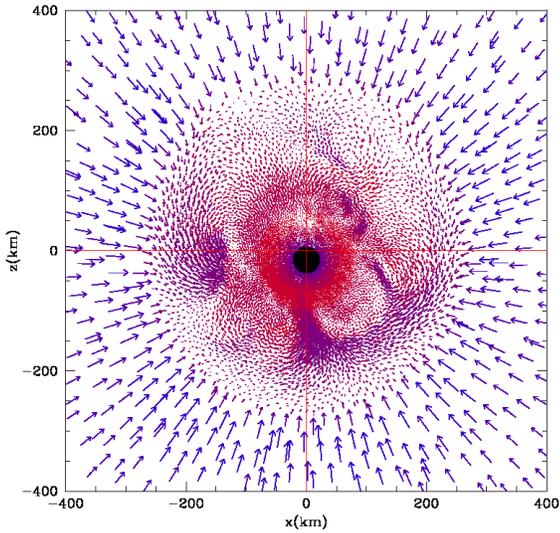}
\caption{A slice of the exploding core of the shell-only 
  simulation 80\,ms after collapse.  Colors denote entropy and vectors
  give velocity direction (vector length denotes velocity magnitude).
  The black dot denotes the core or proto-neutron star (defined, as in
  Fig. 1, by that material with densities above $10^{13} {\rm g\,
    cm^{-3}}$).  The red lines show the x and y axis.  The core of the
  star was initiall centered at x=y=z=0.  Note that the core has
  already moved 20\,km, but that the downflows are strongest on the
  leading edge of the neutron star.}
\end{figure}

\begin{figure}
\plotone{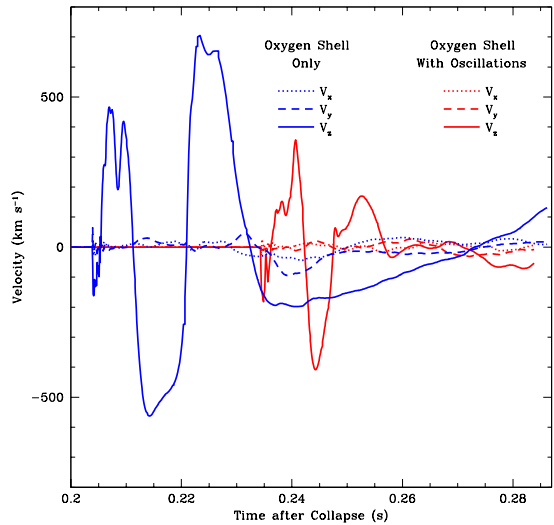}
\caption{x, y, and z velocities versus time for the nascent neutron 
  star for both the shell-only and core oscillation simulations.  The
  oscillatory behavior is due to the neutrino emission from material
  accreting onto the proto-neutron star.  This accretion occurs
  preferentially on the leading edge of the neutron star motion.}
\end{figure}

\begin{figure}
\plotone{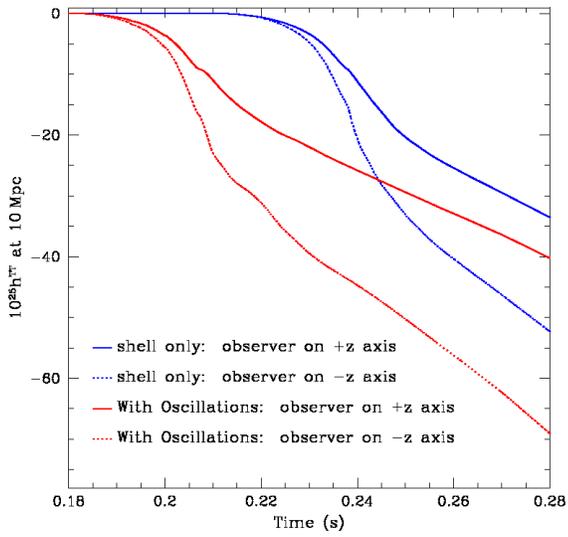}
\caption{Gravitational Wave signal for both the shell only and core 
oscillation simulations as a function of time and observer location.  
Note that the core oscillation simulation has a much stronger 
gravitational wave signal than the signal produced by the shell-only 
simulation.}
\end{figure}

\end{document}